\documentstyle[11pt,newpasp,twoside,epsf]{article} 
\def\edcomment#1{\iffalse\marginpar{\raggedright\sl#1\/}\else\relax\fi}
\reversemarginpar

\begin{document}
\title{Mapping the Galactic Free-Free Foreground via Interstellar
H$\alpha$ Emission}
 \author{R. J. Reynolds and L. M. Haffner}
\affil{Department of Astronomy, University of Wisconsin-Madison,
475 North Charter Street, Madison, WI 53706}

\begin{abstract} 
Recently completed H$\alpha$ surveys of large portions of the sky can be
used to create maps of the free-free intensity distribution at high
Galactic latitude that are independent of the spectral fits to the CMB
data.  This provides an opportunity to test the accuracy of the spectral
fitting procedures and to search for other sources of Galactic forground
contamination that could be confused spectrally with the free-free, such
as spinning dust grains.  The Wisconsin H$\alpha$ Mapper (WHAM) survey has
sampled the sky north of declination $-30\deg$ at about one degree angular
resolution and has revealed that, except for a few isolated regions of
enhanced emission, $\Delta$T$_{ff}$(30 GHz) $\la 30 \, \mu$K at Galactic
latitudes near 15$\deg$, decreasing to $\Delta$T$_{ff}$(30 GHz) $\la 4
\,\mu$K at latitudes above 50$\deg$. Also in progress are H$\alpha$
surveys that sample the sky at higher angular resolution.
\end{abstract}

\section{Introduction}

Diffuse, low density ionized hydrogen at a temperature near $10^4$ K is   
wide spread throughout the disk and halo of the Milky Way (e.g., Reynolds
1991).  The thermal bremsstrahlung associated with this gas is expected to
be one of the principal sources of foreground contamination to the CMB
observations at mm wavelengths (Bennett et al 1992).  However, because
Galactic synchrotron emission dominates the spectrum at lower frequencies 
and thermal emission from dust dominates at higher frequencies, the   
distribution of the free-free emission over the sky cannot be directly
observed.

Fortunately, hydrogen recombination lines, and H$\alpha$ in particular,
provide a way to trace the ionized hydrogen and thus the free-free.  A
thorough review of the implications of H$\alpha$ observations for studies
of the CMB has been provided by McCullough et al (1999).  This paper is a
brief update with an emphasis on the recently completed Wisconsin
H$\alpha$ Mapper (WHAM) survey.

\section{Relationship between H$\alpha$ and Free-Free Emission}

Valls-Gabaud (1998)  has shown that the brightness temperature $T_{ff}$ of
the free-free can be related to the H$\alpha$ intensity by

\begin{equation}
\frac{T_{ff}}{I_{H\alpha}} =
\frac{1.56}{\nu_{30}^2}
\, T_{4}^{0.317} \, 10^{0.029/T_4}
\left(1 + \frac{n_{He~II}}{n_{He}} \,
\frac{n_{He}}{n_H} \right) g_{ff} (\nu ,T) 
\; \; \frac{\mu K}{R}
\end{equation}\\
where 1 R = $10^6/4\pi$ photons s$^{-1}$ cm$^{-2}$ sr$^{-1}$, $\nu_{30}$
is the frequency in units of 30 GHz, $T_4$ is the temperature in units of
10$^4$ K, $n_{He~II}/n_{He}$ is the fraction of helium that is singly
ionized, $n_{He}/n_H$ the abundance of helium, and
$g_{ff}$ is an approximation for the Maxwellian averaged Gaunt factor,
given by

\begin{equation}
g_{ff}(\nu ,T) \approx 4.6 \, T_4^{0.21} \, \nu_{30}^{-0.14}.
\end{equation}\\
This relation between $T_{ff}$ and $I_{H\alpha}$ is valid for frequencies
relevant to the CMB and for sight lines in which the interstellar
extinction of the H$\alpha$ emission is neglectable.  Corrections for
extinction could be problematic because the decrease in the observed
H$\alpha$ surface brightness in a given direction depends not only on the
total column of dust but upon the unknown arrangement of the emitting and
absorbing material along the line of sight.  In addition, because the
source is extended on the sky, the albedo and scattering phase function of
the grains would need to be characterized.  Fortunately, at high Galactic
latitudes the extinction at H$\alpha$ is small.  For example, near the
Galactic poles, E(B-V) $\approx$ 0.018 (Schlegel, Finkbiner, and Davis
1998), corresponding to an optical depth $\tau_{H\alpha} \approx 0.04$; 
$\tau_{H\alpha}$ reaches unity for sight lines within 10$\deg$ of the
midplane.

The ionization state of helium in the emitting gas is also uncertain.
Tufte (1997) and Reynolds and Tufte (1995) found that $n_{He~II}/n_{He}$
was approximately 0.5 or less near the Galactic midplane.  However, Rand
(1997) observed values closer to unity above the midplane of the edge-on
galaxy NGC~891, a galaxy similar to the Milky Way.  Thus it is possible
that along high latitude sight lines in the Milky Way, which sample gas
far from the midplane, $n_{He~II}/n_{He} \approx 1$.  Doubly ionized
helium is assumed to be negligible.

The temperature of the emitting gas has been explored through measurements
of line widths from different mass ions as well as from the intensities of
collisionally excited forbidden lines.  A comparison of the width of the
H$\alpha$ line to that of the [S~II] $\lambda$6716 line indicates a mean
temperature of approximately 8000 K (Reynolds 1985).  Studies of
[S~II] $\lambda$6716/H$\alpha$ and [N~II] $\lambda$6584/H$\alpha$
intensity ratios by Haffner et al (1999) suggest that temperatures vary
from direction to direction, ranging from about 6000 K to 10,000 K or
more.  This would translate into $\pm$11\% variations in
$T_{ff}/I_{H\alpha}$.

Therefore, for $T_4 = 0.8$, $n_{He~II}/n_{He} = 1$, $n_{He}/n_H = 0.08$,
and no correction for extinction,

\begin{equation}
\frac{T_{ff}}{I_{H\alpha}} \approx 7.4 \, \nu_{30}^{-2.14} \; \;
\frac{\mu K}{R}.
\end{equation}

\section{High Galactic Latitude H$\alpha$ Observations}

Gaustad, McCullough, and Van Buren (1996) and Marcelin et al (1998) were
the first to use this relationship between H$\alpha$ and free-free
emission to set upper limits on the free-free contamination of CMB
observations sampling small regions of the northern and southern
hemisphere, respectively. Subsequently, much more extensive H$\alpha$
mapping programs have been undertaken in an effort to explore the
distribution and kinematics of the warm ionized gas in the Galaxy,
particularly at high latitudes.

\subsection{The WHAM Survey}

The Wisconsin H$\alpha$ Mapper (WHAM) survey has provided the first
comprehensive view of the distribution and kinematics of the diffuse
interstellar H$\alpha$ emission at high Galactic latitudes (Haffner et al,
in preparation).  WHAM consists of a 15 cm aperture, dual etalon
Fabry-Perot spectrometer coupled to a dedicated 0.6 m telescope (Tufte
1997). For the sky survey WHAM provided a one degree diameter beam on the
sky and recorded the average spectrum within that beam at 12 km s$^{-1}$
resolution within a 200 km s$^{-1}$ wide spectral window centered on the
H$\alpha$ line.  The tandem Fabry-Perot design provided high spectral
purity by suppressing the multi-order ``ghosts'' arising from the
relatively bright terrestrial OH emission, which combined with the high
spectral resolution, also made it possible to resolve clearly the weak
interstellar H$\alpha$ emission from the much brighter geocoronal
H$\alpha$ line.  The WHAM facility is located at Kitt Peak near Tucson,
Arizona and operated remotely from Madison, Wisconsin.

\begin{figure}

  \plotone{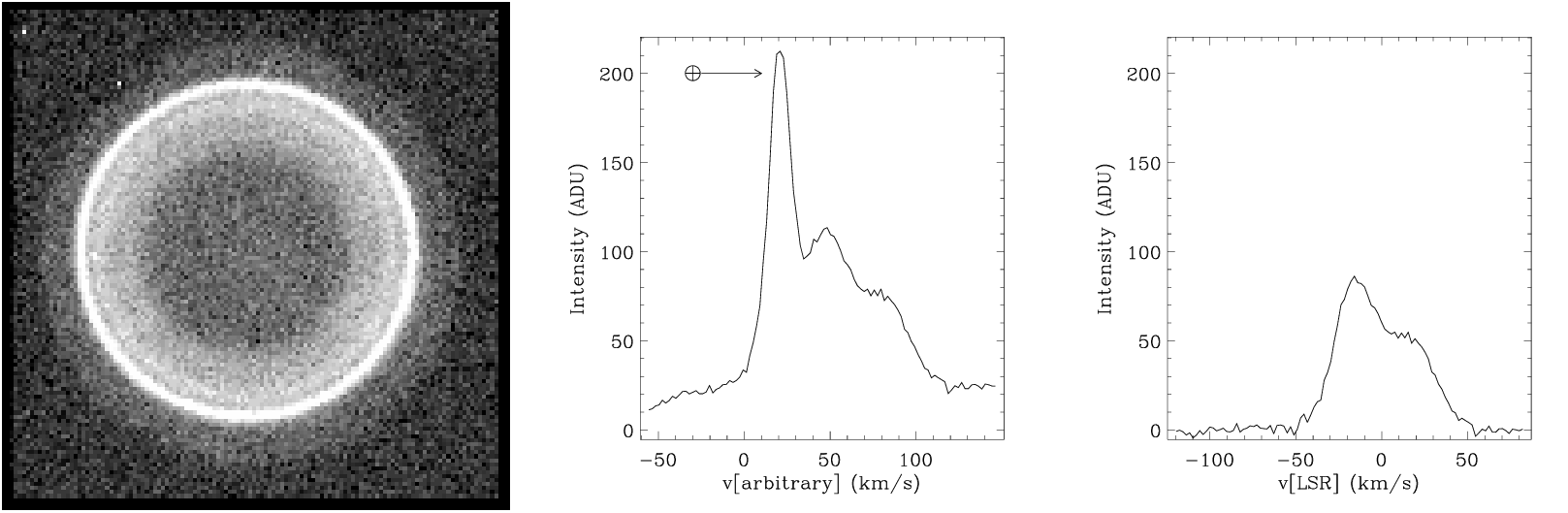}

  \caption{ WHAM H$\alpha$ survey data for $l = 203\fdg8, b = -43\fdg3$.
    a) the raw CCD image (30 s exposure); b) the resulting spectrum
    ($\oplus$ denotes the geocoronal line); c) the pure interstellar
    spectrum after flat fielding, removal of the geocoronal line,
    subtraction of the sky continuum, and registration of the velocity
    scale to the LSR.}

\end{figure}

From 1997 January through 1998 September, WHAM obtained approximately
37,000 spectra with its 1\deg\ diameter beam covering the sky on a $0\fdg
98 \times 0\fdg 85$ grid ($l$,b) north of declination $-30\deg$. The
integration time for each spectrum was 30 s.  Figure 1 illustrates a
sample observation from the survey at $l = 203\fdg8$, $b = -43\fdg3$,
showing both the raw CCD image and resulting H$\alpha$ spectra.  The
geocoronal H$\alpha$ line, produced by solar excitation of atomic hydrogen
in the earth's upper atmosphere, is the thin, bright annulus in the CCD
``ring spectrum''.  This emission appears as a prominent, relatively
narrow line in the center frame of Figure 1.  The interstellar emission is
the broader feature inside the geocoronal ring, appearing in this case to
consist of two blended velocity components at +30 km s$^{-1}$ and +60 km
s$^{-1}$ with respect to the geocoronal line. In general, the separation
between the interstellar emission and the geocoronal line is due to a
combination of the earth's orbital velocity, the sun's peculiar velocity,
and intrinsic motions of the interstellar gas, including Galactic
differential rotation.  The observation for each direction was carried
out at a time of the year that resulted in a significant radial
velocity separation between the earth and the Galactic local standard
of rest (LSR).  The two interstellar components in Figure 1 have
a total intensity of about 6 R, corresponding to a free-free brightness
temperature in this direction of 44 $\mu$K at 30 GHz (see \S 2. above).  
The geocoronal line is removed from the data by fitting each spectrum with
gaussian components and then subtracting from the spectrum the fitted
gaussian associated with the geocorona.  The resulting pure interstellar
spectrum is shown in the third frame of Figure 1.  The absolute intensity
calibration was obtained by comparison with standard astronomical sources
(e.g., Scherb 1981) and is accurate to about 15\%.

In addition to the bright geocoronal line, much fainter atmospheric
emission lines were also removed from each spectrum.  Although the
origin of these extremely weak ($\la 0.1$ R) terrestrial lines is not
known, their wavelengths and relative intensities have been
characterized from long integration spectra obtained over a many month
period (Hausen et al, in preparation).  The effect of the removal of
these lines is shown in Figure 2.

\begin{figure}

  \plotone{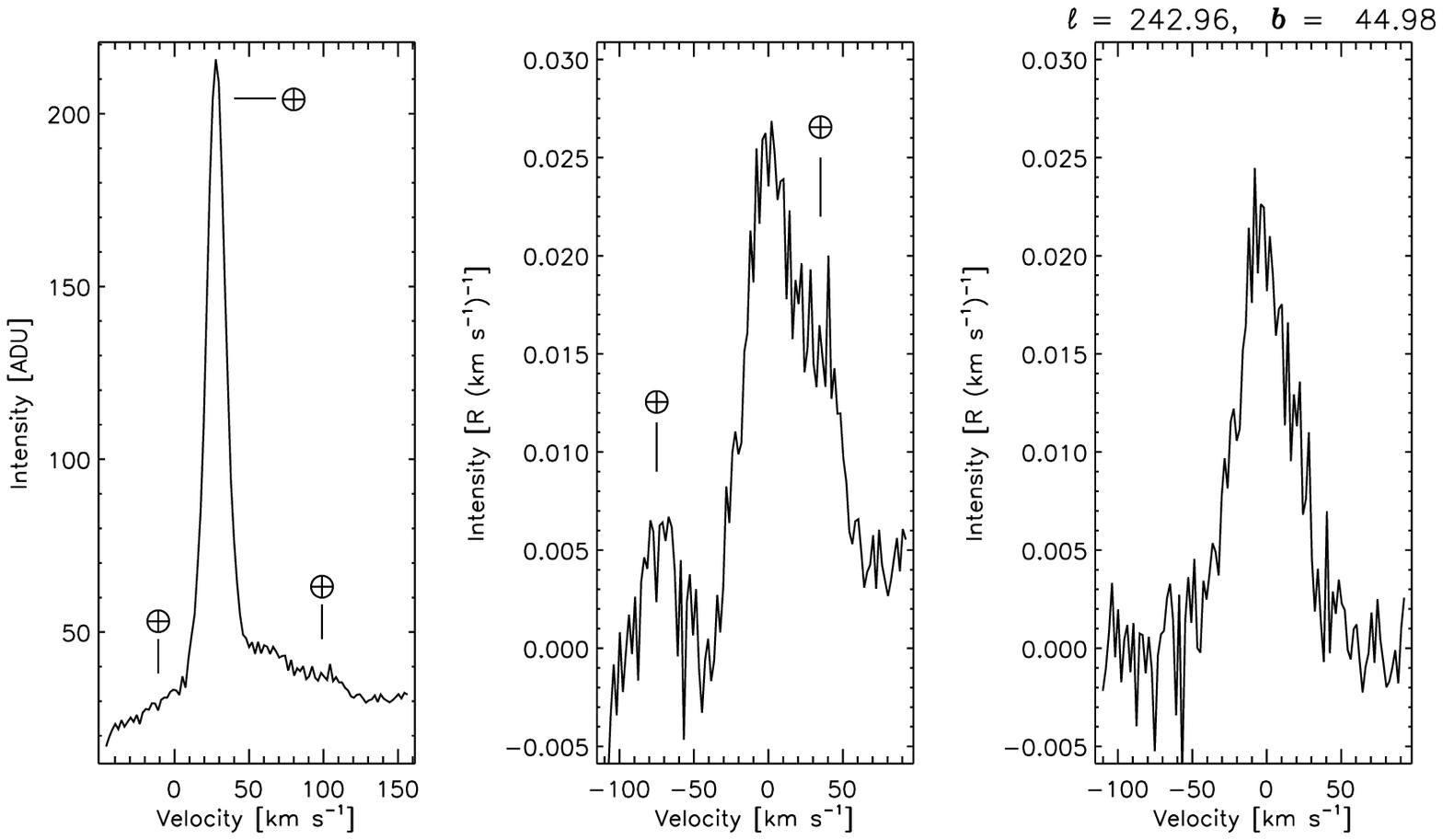}

  \caption{a) Raw WHAM spectrum toward $l = 243\deg$, $b = +45\deg$;
    b) the geocorona H$\alpha$ subtracted spectrum without removal of
    the weaker atmospheric lines ($\oplus$ denotes the most prominent
    atmospheric lines); c) fully processed spectrum. The
    velocity integrated intensity of the Galactic emission shown here
    is about 1 R.}

\end{figure}

All observations were carried out during dark of the moon to avoid
contamination by features in the solar spectrum.  However, zodiacal light
covers the night sky and contains the H$\alpha$ Fraunhoffer absorption
line in its scattered solar spectrum.  No attempt was made to formally
remove this zodiacal feature because there is no clear evidence for it in
the processed spectra (see, for example, the third panel in Fig. 2). Its
apparent absence is due in part to the fact that core of the line is to a
significant degree ``fitted out'' as part of the removal of the geocoronal
line, while the effect of its Lorentzian wings, which are much broader
than the band width of the WHAM spectrum, is just to suppress the baseline
of the spectrum.  Any residual zodiacal absorption feature appears to be
at the level of about $-0.1$ R or less.

The radial velocity interval for the survey was limited to $\pm 100$ km
s$^{-1}$ wrt the LSR.  This range includes nearly all of the interstellar
emission at high latitudes except the H$\alpha$ associated with
High Velocity H~I Clouds (HVCs), which by definition have radial
velocities $|$v$| >$ 80 km s$^{-1}$.  Tufte, Reynolds, \& Haffner (1998)
observed a number of the large high latitude HVC complexes and found that
their H$\alpha$ surfaces brightnesses are within a factor of two of 0.1 R.  
Therefore, the fact that HVCs were excluded from the WHAM survey pass band
implies that in those parts of the sky occupied by HVCs (see Wakker \& van
Woerden 1997), the H$\alpha$ intensity on the WHAM survey is too low by
about 0.1 R (0.7 $\mu$K at 30 GHz).

Figure 3 shows a portion of the beam covering pattern for the sky
survey.  The observations were obtained in ``blocks'', with each block
usually consisting of 49 pointings made sequentially in a
boustrophedonic raster of seven beams in longitude and seven beams in
latitude.  Each block took approximately 30 minutes, and from one to
twenty blocks were observed in a night.  The absence or presence of
block boundary features in the completed survey map provide an excellent
test of the systematic errors associated with observations taken on
different nights and different times of the year.

\begin{figure}

  \plotone{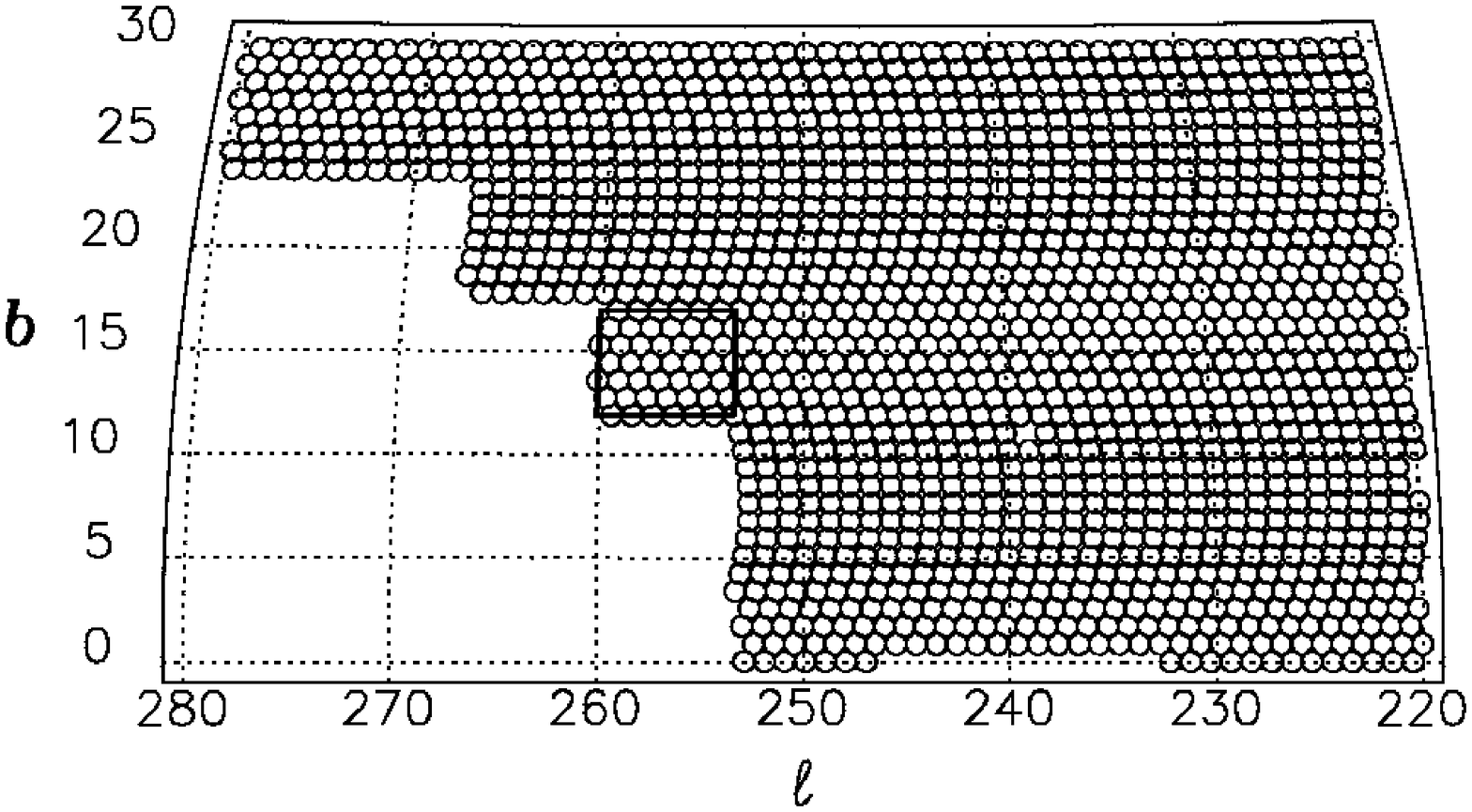}

  
  \caption{A portion of the sky showing the pattern of WHAM's 1$\deg$
    diameter beams. Observations were obtained as a sequence of
    ``blocks'' (outlined), consisting typically of a grid of 49
    pointings within in a $7\deg \times 6\deg$ region. The integration
    time per pointing was 30 s.}

\end{figure}

Figure 4 is the resulting survey map showing the total H$\alpha$ intensity
within the $\pm 100$ km s$^{-1}$ interval centered near the local standard
of rest.  Interstellar H$\alpha$ emission extents over virtually the
entire sky, with blobs and filaments of enhanced H$\alpha$ superposed on a
more diffuse background.  The highest H$\alpha$ intensities are found near
the Galactic equator, with a general decrease toward the poles.  The
relatively bright high latitude enhancement at $l = 315\deg$, $b =
+50\deg$ is the H~II region ionized by Spica (B1~III-IV + B2~V), and some
of the other high latitude enhancements also seem to be associated with
either B stars or hot white dwarf stars, such as $\pi$ Aqr (B1~V) at
66$\deg$, $-45\deg$ and sdO PHL 6783 at 124$\deg$, $-74\deg$ (see Haffner
2000). The large, relatively bright region of diffuse and filamentary
emission located between longitudes 180$\deg$ and 240 $\deg$ and latitudes
$-5\deg$ and $-50\deg$ is the Orion-Eridanus bubble (Heiles, Haffner, \&
Reynolds 1999).

\begin{figure}


  \plotone{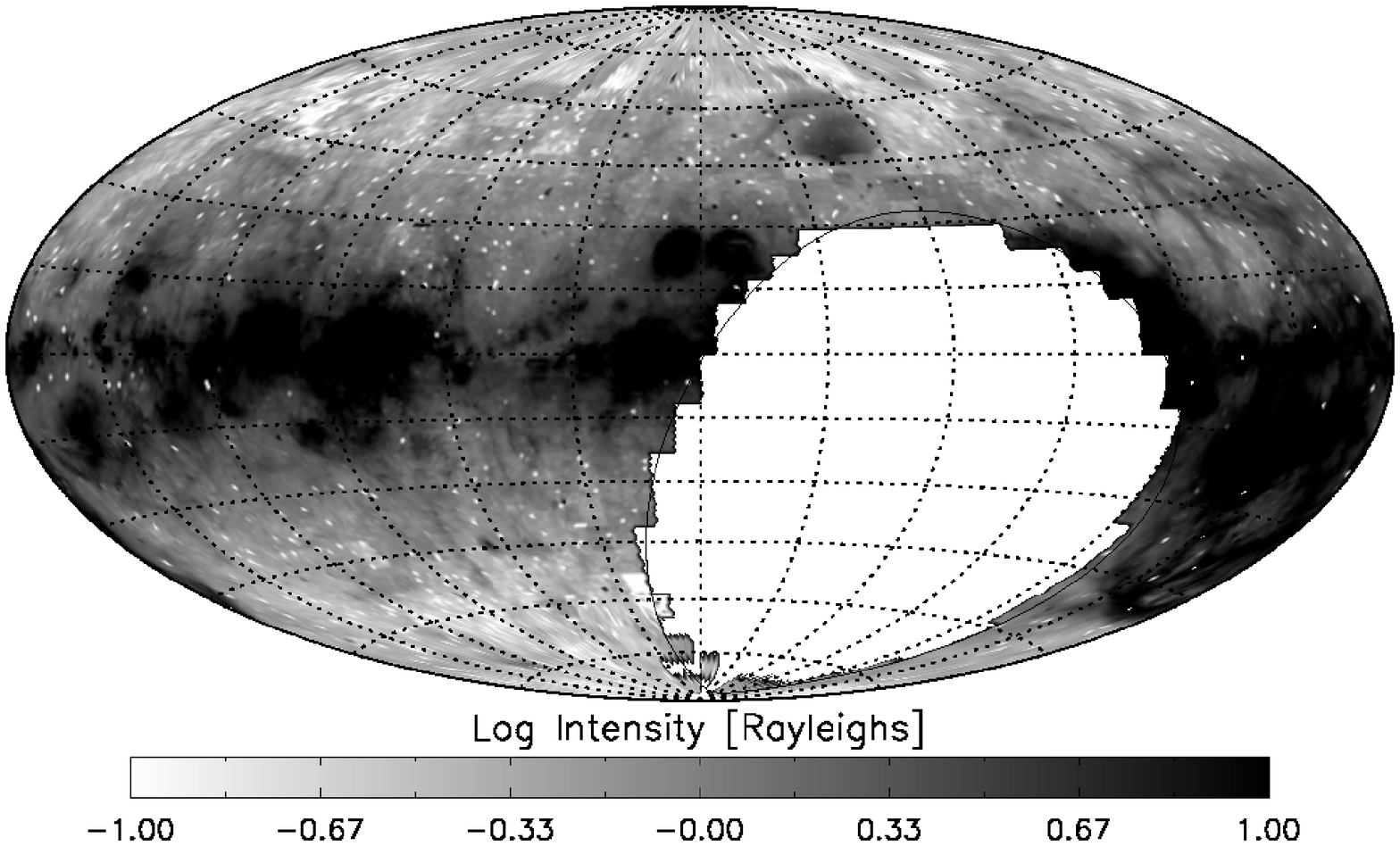}

  \caption{The WHAM H$\alpha$ Sky Survey. The total (velocity-integrated)
    H$\alpha$ intensities for over 37,000 observations are plotted on
    a Hammer-Aitoff projection centered at $l = 0\deg$, $b = 0\deg$.
    The gray scale represents the intensity within each
    1$\deg$-diameter beam.  In this display the intensity scale is
    limited to the range 0.1 R to 10 R (0.7 $\mu$ K to 74 $\mu$ K at
    30 GHz). Dotted grid lines are separated by 30$\deg$ in Galactic
    longitude and 15$\deg$ in latitude.  The solid line denotes the
    $-30\deg$ declination limit of the survey.  The white dots scattered
    across the map are beams containing bright (V $\la$ 6 mag) stars with
    a significant Fraunhoffer absorption line at H$\alpha$.}

\end{figure}

The distribution of H$\alpha$ intensities vs Galactic latitude is shown in
Figure 5.  This plot shows that except for a few isolated regions of 
enhanced emission, the H$\alpha$ intensity variations on scales of one 
degree or larger are generally less than 4 R (30 $\mu$K at 30 GHz) at
$|$b$| \approx 15\deg$, $\la$ 1.7 R (13 $\mu$K) at $|$b$| \approx  
30\deg$, and $\la$ 0.5 R (4 $\mu$K) at latitudes higher than 50$^{\rm
o}$.

\begin{figure}


  \plotfiddle{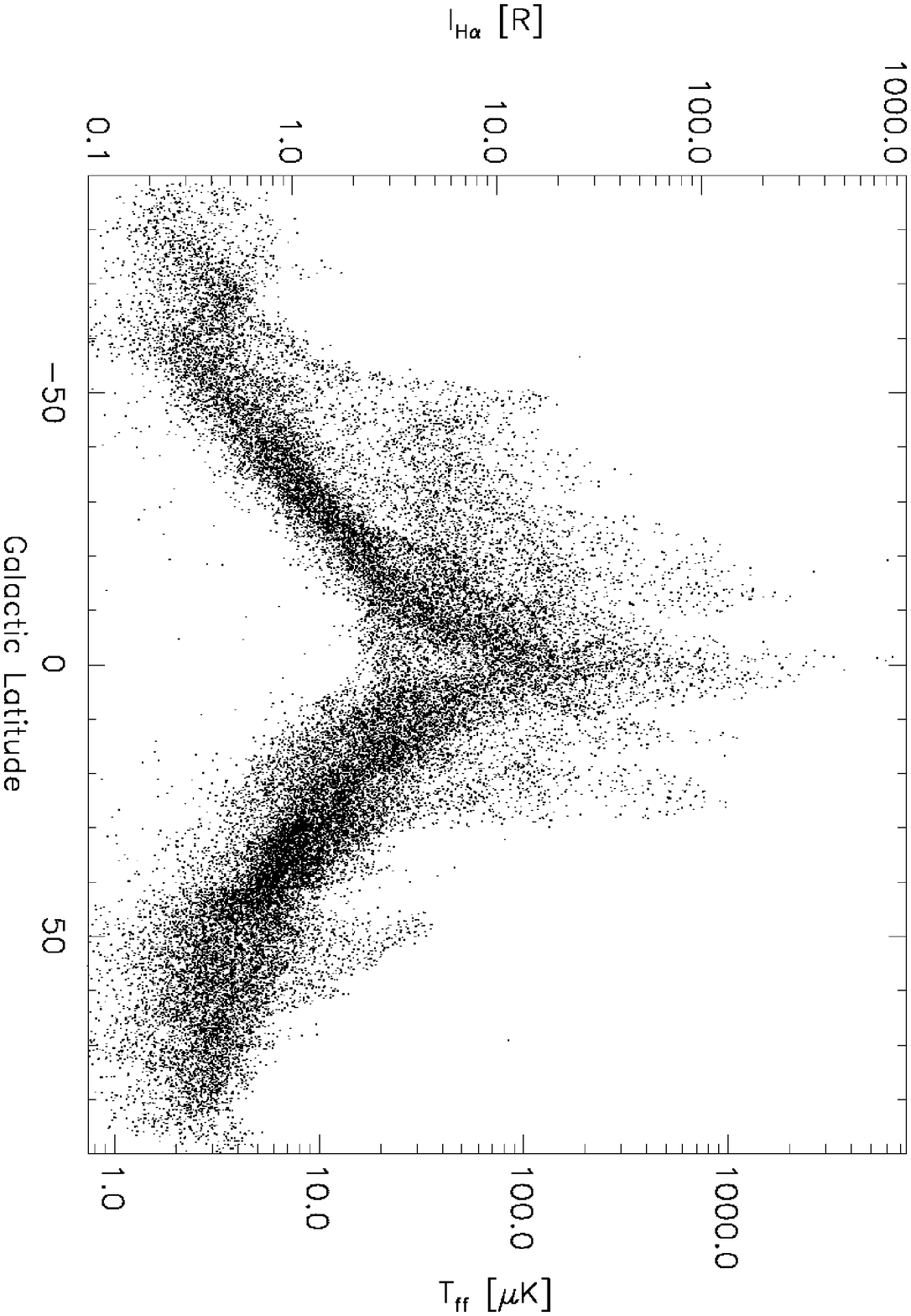}{3in}{90}{50}{50}{200}{-40}

  \caption{The H$\alpha$ intensities from the WHAM survey and
    corresponding values for T$_{ff}$ at 30 GHz from eq. (3) are
    plotted vs Galactic latitude.  Within 10\deg\ of the midplane the
    H$\alpha$ intensities are limited significantly by interstellar
    extinction.}

\end{figure}

\subsection{High Spatial Resolution (Sub-Degree) Imaging Surveys}

Wide-field CCD imaging has been used to explore the interstellar H$\alpha$
on angular scales significantly smaller than 1 degree.  This has been
accomplished at the expense of spectral resolution, making it impossible
to remove the geocoronal H$\alpha$, OH, and other atmospheric emissions.
Therefore, the imaging and WHAM spectral surveys complement each other,
with the imaging providing valuable new information about the smaller
scale structure and WHAM providing the absolute intensity scale and
sensitivity to faint, larger scale emission features.

An imaging survey of the sky south of declination $+15\deg$ has
been undertaken by Gaustad et al (1998).  The observations, made with a
small robotic CCD camera at Cerro Tololo Inter-american Observatory in
Chile, consist of 269 $13\deg \times 13\deg$ images with centers
the same as those in the IRAS Sky Survey Atlas.  On-band images are
obtained through a 30 \AA\ wide H$\alpha$ filter with corresponding
off-band images obtained through a dual-pass filter that transmitted two
60 \AA\ wide bands on either side of H$\alpha$.  The resulting on-band
minus off-band images, median filtered to $5\arcmin \times 5\arcmin$, show
interstellar H$\alpha$ emission features down to about 1 R with excellent
subtraction of the stellar images (see McCullough et al 1999).  Most of
the observations have been completed, and reduction is underway. Figure 6
compares portions of the Gaustad et al and WHAM surveys.  A similar
imaging survey, confined to declinations north of $-15\deg$ and
Galactic latitudes $|$b$| < 30\deg$, is being carried out by
Dennison, Simonetti, \& Topasna (1998) from Virginia, USA.

\begin{figure}

  \plotone{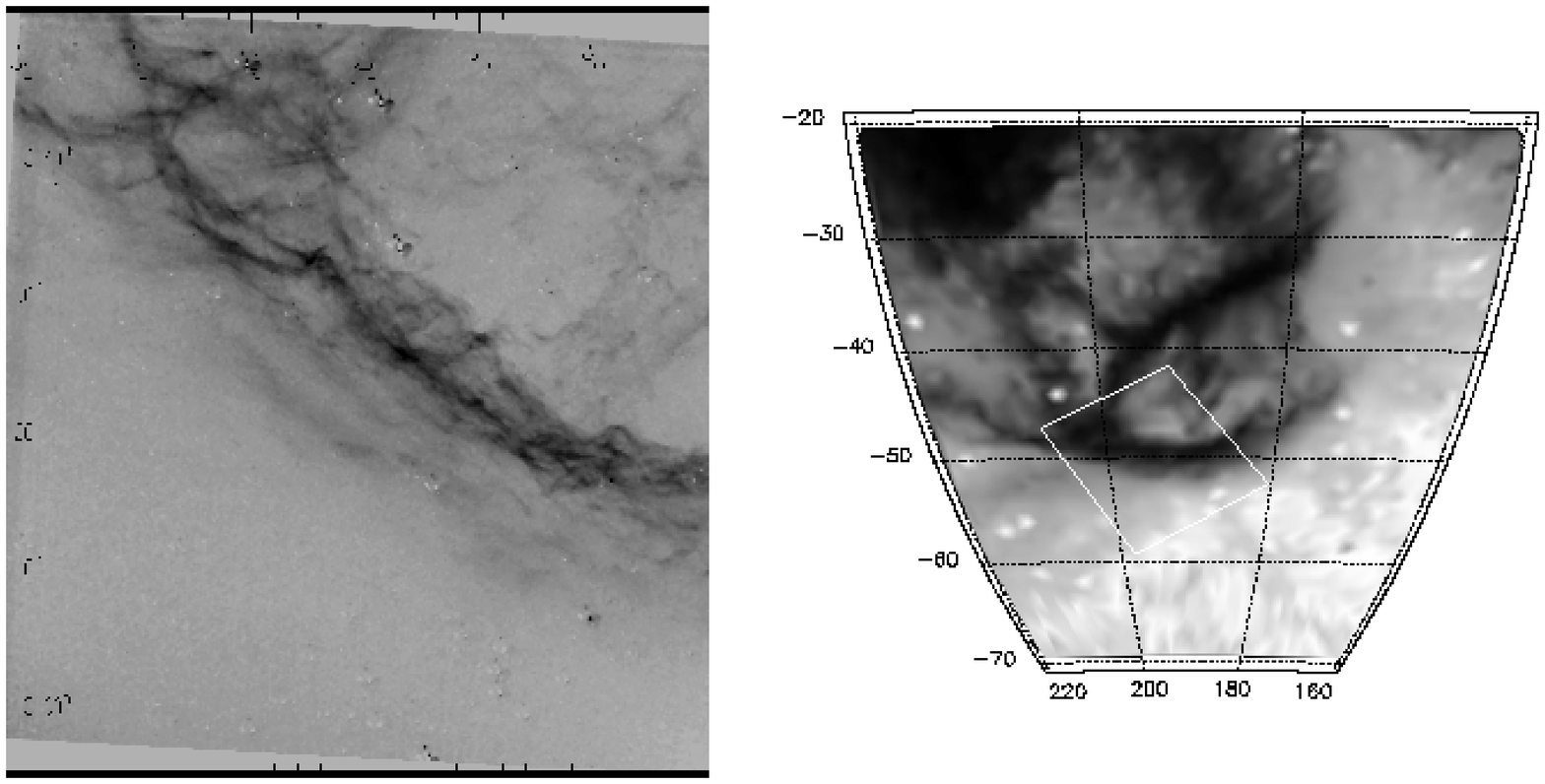}


  \caption{A $12\deg \times 12\deg$ region from the Gaustad et al
    imaging survey (left) shows detailed structure in the high latitude
    H$\alpha$ filament associated with the Orion-Eridanus bubble. The
    corresponding portion of the WHAM survey is also shown (right) with
    the same region of the sky as the high resolution image outlined in
    white.  Note that the high resolution image is rotated about
    30\deg clockwise with respect to the WHAM map.}

\end{figure}

\section{Comparing H$\alpha$ with Surveys at Other Wavelengths}

Because the WHAM data have just been reduced and the reduction of the
high latitude imaging is still in progress, there has not yet been an
opportunity to carry out formal analyses incorporating large portions of
the sky, to search for correlations between H$\alpha$ and the H~I, dust,
and synchrotron emissions, for example.  On the WHAM survey map (Fig.
4), except for the csc$|$b$|$ law, no striking correspondences are
readily apparent by eye. The H$\alpha$ features in the Orion-Eridanus
region do have some correspondence with the 21 cm, dust, and soft X-ray
emission, presumably associated with energetic events in Orion OB1
(Heiles et al 1999; Reynolds \& Ogden 1979); however, other large scale
features, like the filament extending from the midplane to $b = +50\deg$
at $l = 224\deg$, seem to have no relationship to nearby hot stars or
the other phases of the interstellar medium (Haffner, Reynolds, \& Tufte
1998).  Studies of small parts of the H$\alpha$ sky have suggested that
some ($\approx$ 30\%) of the H$\alpha$ is associated with H~I regions
(Reynolds et al 1995), and that the amount of thermal dust emission
associated with the diffuse H~II varies significantly across the sky
(Kogut 1997; Arendt et al 1998; McCullough et al 1999; Lagache et al
2000; de Oliveira-Costa et al 2000).  A comparision of WHAM data with
the 26-46 GHz QMAP observations near the North Celestial Pole shows a
correlation with the H$\alpha$ consistent with that predicted by
equation 3 (de Oliveira-Costa et al 2000), although with marginal
significance.  More comprehensive analyses using much larger portions of
the sky will hopefully further clarify these results.  Also, comparisons
of the free-free emission, as revealed by the H$\alpha$, with the total
``free-free'' foreground intensity derived from the spectral fits to the
CMB data should help to quantify the amount of microwave emission
produced by spinning dust grains (Draine \& Lazarian 1998; Ferrara \&
Dettmar 1994).

\section{Summary and Conclusions}

Thermal bremsstrahlung from the warm ionized component of the interstellar
medium is one of the principal Galactic foregrounds contaminating
observations of the cosmic microwave background (CMB).  Near the peak of
the CMB, the intensity of the free-free may be comparable to that of the
Galactic synchrotron and thermal dust emission; however, because it does
not dominate the light from the sky at any wavelength, its angular
distribution over the sky cannot be observed directly.  Within the last
few years, surveys of diffuse interstellar H$\alpha$ emission have been
undertaken to explore the warm ionized interstellar gas.  At high Galactic
latitudes, where interstellar extinction of the optical H$\alpha$ is
small, the relationship between the intensity of the H-recombination line
and the free-free continuum at a given frequency is only a function of the
electron temperature. Since the temperature of the gas is tightly
constrained, these high latitude H$\alpha$ surveys can provide accurate
maps of the Galactic free-free emission, and thus play a potentially
important role in testing the accuracy of the spectral fits to CMB
foregrounds and exploring the degree to which spinning dust grains
contribute to the foreground emission.

\section{Acknowledgements}

We would like to acknowledge Steve Tufte, Nikki Hausen, Brian Babler, Greg
Madsen, Rebecca Pifer, Mark Quigley, Kurt Jaehnig, and Jeff Percival for
their work on the WHAM survey. We also thank Peter McCullough for
providing a sample from their imaging data.  WHAM is supported by the
National Science Foundation through grant AST 96-19424.

\end{document}